\documentclass[12pt]{article}

\usepackage{scicite}
\usepackage{graphicx}
\usepackage{amsmath,nicefrac,amssymb}

\usepackage{times}

\newenvironment{sciabstract}{%
\begin{quote} \bf}
{\end{quote}}

\newcommand{\AM}[0]{\mbox{AM }}

\newcounter{lastnote}

\title{Attosecond Coherent Electron Motion in
Auger-Meitner Decay}

\author
{Siqi Li,$^{1,2\ast}$ Taran Driver,$^{1,3,4\ast}$ Philipp Rosenberger,$^{1,3,5,6}$ Elio G. Champenois,$^{3}$ \\ Joseph Duris,$^{1}$ Andre Al-Haddad,$^{7}$ Vitali Averbukh,$^{4}$ Jonathan C. T. Barnard,$^{4}$\\Nora Berrah,$^{8}$
Christoph Bostedt,$^{7,9}$ Philip H. Bucksbaum,$^{2,3,10}$ Ryan Coffee,$^{1,3}$ \\ Louis F. DiMauro,$^{11}$ Li Fang,$^{11,12}$ Douglas Garratt,$^{4}$ Averell Gatton,$^{1}$ \\ Zhaoheng Guo,$^{1,10}$ Gregor Hartmann,$^{13}$ Daniel Haxton,$^{14}$ Wolfram Helml,$^{15}$ \\ Zhirong Huang,$^{1,2}$ Aaron C. LaForge,$^{8}$ Andrei Kamalov,$^{1,3,10}$ \\ Jonas Knurr,$^{3}$ Ming-Fu Lin,$^{1}$ Alberto A. Lutman,$^{1}$ James P. MacArthur,$^{1,2}$ \\ Jon P. Marangos,$^{4}$ Megan Nantel,$^{1,2}$ Adi Natan,$^{3}$ Razib Obaid,$^{8}$ \\ Jordan T. O'Neal,$^{2,3}$ Niranjan H. Shivaram,$^{1}$ Aviad Schori,$^{3}$ Peter Walter,$^{1}$\\ Anna Li Wang,$^{3,10}$ Thomas J. A. Wolf,$^{1,3}$ Zhen Zhang,$^{1}$ Matthias F. Kling,$^{1,3,5,6}$ \\ Agostino Marinelli,$^{1,3,\dagger}$ James P. Cryan,$^{1,3,\ddagger}$\\
\\
\normalsize{$^{1}$SLAC National Accelerator Laboratory}\\
\normalsize{$^{2}$Department of Physics, Stanford University}\\
\normalsize{$^{3}$Stanford PULSE Institute, SLAC National Accelerator Laboratory}\\
\normalsize{$^{4}$The Blackett Laboratory, Department of Physics, Imperial College
London}\\
\normalsize{$^{5}$Max Planck Institute of Quantum Optics, Garching, Germany}\\
\normalsize{$^{6}$Physics Department, Ludwig-Maximilians-Universit\"at Munich, Garching, Germany}\\
\normalsize{$^{7}$Paul Scherrer Institute}\\
\normalsize{$^{8}$Physics Department, University of Connecticut}\\
\normalsize{$^{9}$LUXS Laboratory for Ultrafast X-ray Sciences, Ecole Polytechnique F\'ed\'erale de Lausanne}\\
\normalsize{$^{10}$Department of Applied Physics, Stanford University}\\
\normalsize{$^{11}$Department of Physics, The Ohio State University}\\
\normalsize{$^{12}$Department of Physics, University of Central Florida}\\
\normalsize{$^{13}$Institut f\"ur Physik und CINSaT, Universit¨at Kassel}\\
\normalsize{$^{14}$KLA Corporation}\\
\normalsize{$^{15}$Zentrum f\"ur Synchrotronstrahlung, Technische Universit\"at Dortmund}\\
\\
\normalsize{$^\ast$These authors contributed equally.}\\
\normalsize{$^\dagger$To whom correspondence should be addressed; E-mail:  marinelli@slac.stanford.edu.}\\
\normalsize{$^\ddagger$To whom correspondence should be addressed; E-mail:  jcryan@slac.stanford.edu.}\\
}

\date{}

\begin{document} 


\baselineskip24pt


\maketitle


\begin{sciabstract}
In quantum systems, coherent superpositions of electronic states evolve on ultrafast timescales~(few femtosecond to attosecond, 1 as = 0.001 fs = 10$^{-18}$ s), leading to a time dependent charge density.  
Here we exploit the first attosecond soft x-ray pulses produced by an x-ray free-electron laser to induce a coherent core-hole excitation in nitric oxide.
Using an additional circularly polarized infrared laser pulse we create a clock to time-resolve the electron dynamics, and demonstrate control of the coherent electron motion by tuning the photon energy of the x-ray pulse.
Core-excited states offer a fundamental test bed for studying coherent electron dynamics in highly excited and strongly correlated matter.
\end{sciabstract}

\paragraph*{Introduction}

Interference is a pillar of quantum physics, and a manifestation of one of its most remarkable consequences: the wavelike nature of matter.
A quantum system can exist in a superposition of energy states whose quantum phases progress to interfere constructively or destructively as the system evolves, causing physical observables (e.g. charge density) to oscillate in time. 
Such oscillations are known as quantum beats~(QBs), and have a period of $T_{QB}=\nicefrac{h}{\Delta E}$, where $h$ is Planck's constant and $\Delta E$ is the energetic separation between the states \cite{forrester1955photoelectric,alexandrov_interference_1964,hadeishi_direct_1965,mauritsson2010attosecond,goulielmakis_real-time_2010}.
In order to display a quantum beat, two conditions must be satisfied: First, the quantum system must be prepared in a superposition of two or more different energy states that have a well-defined~(or coherent) relationship between their individual quantum phases, and that this relationship stays stable during the characteristic time of these energy states.
Second, the physical observable must be sensitive to the overall quantum phase of this coherent superposition. 

In this work, we demonstrate the creation and observation of coherent superpositions of core-excited states in molecules using attosecond x-ray pulses.
These molecules decayed non-radiatively \textit{via} the Auger-Meitner~(\AM) mechanism; a multi-electron process where the core vacancy created by an x-ray pulse is filled by one electron from a valence orbital, while another valence electron is emitted to conserve energy. 
The \AM process is the dominant mechanism for relaxation following x-ray absorption in most biologically relevant molecules, and any molecules composed of light atoms such as carbon, oxygen, and nitrogen. 

We sought to learn how coherence in short X-ray pulses is imprinted on excited electronic states following x-ray/matter interaction, and how this affects the attosecond evolution of the excited electronic wavepacket.
To this end, we measure the time-dependent \AM yield, and find that it is sensitive to the quantum coherence of the electronic wavepacket, as well as the differences in the excited state populations.
The coherence of the wavepacket is manifested as femtosecond modulations~(or quantum beats) in the time-dependent electron yield.
This coherence in the relaxation process could affect a broad class of other ultrafast measurements where photoabsorption must be taken into account, from protein crystallography to x-ray photochemistry,
because the need for high temporal resolution necessitates the use of broad bandwidth X-ray pulses.  

Time-resolved measurements of any correlated electron interaction~(including \AM decay) are challenging due to the extreme timescale~(few- to sub-femtosecond) on which electron-electron interactions occur.
Previous time-resolved measurements have extracted a single parameter~($\Gamma$) to characterize the decay of a core-excited system~\cite{drescher_time-resolved_2002,uiberacker_attosecond_2007,verhoef_time-and-energy_2011,haynes2021clocking}.
In the case of short excitation/ionization pulses, $\Gamma$ corresponds to the lifetime of the core-excited state, but for long pulses the extracted decay constant is altered by interferences with the excitation process~\cite{haynes2021clocking,smirnova_quantum_2003,kazansky_time-dependent_2009}.
Our unique combination of short excitation pulses and a sufficiently long observation window allows for a direct time-resolved measurement of the \AM emission process.   
We measure a quantum beat, demonstrating the creation and observation of electronic coherence in a core-excited molecular system.
Our technique of mapping coherent electronic motion to the \AM decay profile offers a unique test-bed for studies of electronic coherence in highly excited and strongly correlated systems.

\paragraph*{Measurement}

The experimental setup used is shown in panel~\textbf{a} of Figure~1.
Isolated soft x-ray attosecond pulses from a free-electron laser~\cite{duris_tunable_2020}, tuned near the oxygen 1$s\rightarrow\pi$ resonance in nitric oxide~(NO) ($\sim$530--540~eV), irradiate a gas target in the presence of a circularly polarized 2.3~$\mu$m laser field.
The momentum distribution of the resultant photoelectrons is recorded by a co-axial velocity map imaging spectrometer~(c-VMI)~\cite{li_co-axial_2018}.
Interaction with the x-ray pulse produces electrons from several different photoionization channels: direct ionization of nitrogen $K$-shell electrons, $KLL$ \AM emission resulting from the nitrogen $K$-shell vacancy, and resonant oxygen \AM emission following O 1$s\rightarrow\pi$ excitation.
These channels are labeled in panel \textbf{b} of Fig.~1, which shows the electron momentum distribution recorded without the 2.3~$\mu$m laser field.
The 1$s\rightarrow\pi$ excitation in nitric oxide corresponds to the promotion of an oxygen $1s$ electron to the degenerate $2\pi$ molecular orbital, which is already partially occupied by an unpaired valence electron.
The resonant \AM emission following this excitation has a dominant feature corresponding to channels where one of the degenerate $2\pi$ electrons participates in the decay, leading to excited cationic states.
There is a small contribution from the channel where both $2\pi$ electrons participate, resulting in a $2\pi^0$ ground configuration of the cation~\cite{wang_is_2003}.

The circularly polarized laser field maps the temporal profile of the electron emission on to the momentum measured at the detector. 
When electrons are released from the molecule following interaction with the x-ray pulse, their trajectory is altered by the presence of the infrared laser field, similar to the principle of a time-resolving streak camera~\cite{bradley1971direct,tsuchiya_advances_1984}.
This interaction changes (or `streaks') the final electron momentum, which is measured at the detector.
In a semi-classical approximation, the final momentum of an ionized electron is given by
\begin{equation}
    \vec{p}(t\rightarrow\infty) = \vec{p}_0 - \vec{A}(t_0),
    \label{Eqn:StreakMomentum}
\end{equation}
where $\vec{A}(t_0)=\int_{-\infty}^{t_0}\vec{\mathcal{E}}_L(t^{\prime}) dt^{\prime}$ is the vector potential of the circularly polarized laser field,~$\mathcal{E}_L(t)$, at the time of ionization~$t_0$, and $\vec{p}_0$ is the momentum of the electron in the absence of the infrared laser field. All the quantities are expressed in atomic units.

In our measurement the pulse duration of the circularly polarized `streaking' laser field~($\sim100$~fs) is much longer than the laser period~($T_L=7.7$~fs).
This implies that over the timescale of the electron emission process the vector potential has nearly constant amplitude~($|\vec{A}|$) but a direction that rotates with constant angular velocity $\nicefrac{2\pi}{T}$.
Thus, Eqn.~\ref{Eqn:StreakMomentum} describes how the streaking technique encodes the temporal evolution of the electron emission rate onto the electron momentum spectrum: an electron emitted at $t_i$ will experience a momentum shift in the direction of $-\vec{A}(t_i)$.
Because the period $T_L$ of the circularly polarized laser is well known, a change in streaking direction of $\Delta\theta$ straightforwardly maps to a change in emission time $\Delta\tau$ by:
\begin{equation}
\Delta\tau=\frac{\Delta\theta}{2\pi}\times T.
\label{eqn:angletotime}
\end{equation}
This mapping of angle-to-time resembles the face of a clock, which has led to the term `attoclock' being used to describe this type of time-resolved measurement~\cite{itatani_attosecond_2002,hartmann_attosecond_2018,eckle_attosecond_2008}.

Our method for extracting the temporal profile of the \AM electron yield is illustrated in Fig.~1, panels~\textbf{d} and ~\textbf{e}.
The synchronization of the streaking laser and x-ray pulse has a jitter of roughly $\sim500$~fs~\cite{glownia_time-resolved_2010}, which is orders of magnitude below the required precision for directly timing the \AM process.
We address this using a single-shot diagnostic of the relative arrival time between the x-rays and laser pulse. 
In addition to driving resonant excitation near the oxygen $K$-edge, the attosecond x-ray pulse ionizes electrons from the nitrogen $K$-shell of the NO molecule (see Fig.~1, panels \textbf{b} and~\textbf{c}).
This direct photoionization process produces high energy ($\sim120$~eV) electrons. 
The photoionization delay between the arrival of the x-ray pulse and the appearance of these fast photoelectrons in the continuum is negligibly small ($\lesssim$ 5~as) compared to the streaking laser period $T_L$ of 7.7~fs~\cite{dahlstrom_introduction_2012,dahlstrom_theory_2013,serov_interpretation_2013}.
Therefore the momentum shift observed for the nitrogen $K$-shell photoemission feature provides an accurate, single-shot measurement of the direction of the streaking laser vector potential $\vec{-A_0}$ at the time of arrival of the x-ray pulse.

Panel~\textbf{d} of Fig.~1 illustrates the change in differential electron signal induced by the streaking field for shots with three different x-ray arrival times.
To extract the time-dependent emission rate of resonant \AM electrons we monitor a small angular region on the detector~(black wedge) for different x-ray arrival times.
This region is chosen to be slightly higher in momentum than the field-free resonant emission spectrum shown in panel~\textbf{b} of Fig.~1.
This ensures that the \AM electrons recorded at this part on the detector interacted with both the IR and X-ray pulses. 
It is important to monitor a small angular region because electronic interference effects are symmetry-forbidden in the core-excited state of NO, so any interference effects should not appear in an angularly integrated measurement~\cite{wang_is_2003,demekhin_strong_2010}.
We note that the period of the streaking field is chosen to be longer than the dominant timescale of the \AM process
This simplifies interpretation of the streaking measurement by limiting the effect of `wrapping', where electrons released into the continuum at time $\tau$ and $\tau+T_L$ experience a similar momentum kick from the streaking field. 

The time-dependent electron yield is shown in panel~\textbf{e} of Fig.~1 and shows a maximum at $\tau$=0, when $\vec{A_0}$ is directed along the detection direction, and the the core-excited population is at a maximum. 
In addition to an exponentially decaying electron emission rate, we observe a revival in the time-dependent emission rate at $\tau$=3.5 fs. 


\paragraph*{Model}

We model our measurement according to the theory of attosecond streaking of multiple Fano resonances described by Wickenhauser~\textit{et al.}~\cite{wickenhauser_attosecond_2006,wickenhauser_time_2005}.
Our model, illustrated in Fig. 2, includes a ground state which is resonantly coupled to three bound states.
These bound states are also coupled to a single, structure-less continuum, which is dressed by the circularly polarized, $2.3~\mu$m streaking laser field.
The coupling between the bound and continuum states is the result of electron correlation interactions, and drives the \AM decay process. 
The bound states have excitation energies of $531.5$~eV ($^2\Sigma^-$), $532.6$~eV ($^2\Delta$) and $533.5$~eV ($^2\Sigma^+$), which represents the core-excitation spectrum of nitric oxide~\cite{puttner_vibrationally_1999}.
The continuum coupling constant ($\Gamma=170$~meV) is consistent with previous x-ray absorption measurements~\cite{puttner_vibrationally_1999}.
The relative amplitude between transitions to the bound, core-excited states and direct photoionization of valence electrons to the continuum is represented by the Fano parameter, $q_i$~(see SM)~\cite{fano_effects_1961}.
We choose the value for $q_i$ according to the measured absorption spectrum of NO \cite{puttner_vibrationally_1999}. 

The coherent bandwidth of the exciting x-ray source is $\sim5$~eV \cite{duris_tunable_2020}, which is sufficient to span all three bound states.
Symmetry constraints do not allow for the coherent population of the $^2\Sigma^-$ and $^2\Sigma^+$ states \cite{wang_is_2003}, which is excluded from the model.
In both simulation and experiment we tune the central wavelength of the x-ray source across the $1s\rightarrow2p\pi$ resonance (red, green and blue shaded curves, bandwidth drawn to scale with energetic separation of core-excited states).

In simulation we can also calculate the energy-resolved continuum wavefunction in the absence of the streaking-field, shown in panel~\textbf{c} as the build-up of resonant features.
The rate of electron emission (integrated over electron kinetic energy) is shown in panel~\textbf{d}), and we clearly observe an oscillatory emission rate.
Finally, in panel~\textbf{e} we show the population of each core-excited state as a function of time, which again shows oscillatory behavior.

The periodic modulation of the electron emission rate results from the coherent population of the two pairs of excited states $^2\Sigma^-$ \& $^2\Delta$ and $^2\Delta$ \& $^2\Sigma^+$.
Electronic coherence between the pairs of excited states results in consecutive minima~(maxima) in core-excited population due to destructive~(constructive) interference, which manifest in a modulation to the decaying electron emission rate.
Because the core-excited wavepacket consists of states with different angular momentum projections along the molecular axis, the excited state wavepacket produces an excited electron density which `rotates' around the molecular axis, as shown in panel \textbf{f} of Fig.~2.


\paragraph*{Results}

We directly model our experimental observable by computing the asymptotic~($t\rightarrow\infty$) momentum distribution of ionized electrons within the Strong-Field Approximation~(SFA)~\cite{wickenhauser_time_2005} (Fig.~2~\textbf{b}) and performing the same analysis routine as we apply to the experimental data. 
The asymmetry parameters describing emission from the oxygen $1s \rightarrow ^2\Sigma^-$, O $1s \rightarrow ^2\Delta$ and O $1s \rightarrow ^2\Sigma^+$ excitations are expected to be different for each of the three electronic states~\cite{demekhin_strong_2010} and have not previously been measured, meaning the contribution of each channel to emission in the direction of our observation window is not well defined. 
We fit the simulation to the experimental data using the lower kinetic energy limits of the small detector region defined in Fig. 1~\textbf{e}, and the relative contribution from each decay channel at the precise region on the detector, as free parameters.
Further details are provided in the Supplementary Materials.
We also account for the possibility of a small systematic error in $t_0$ determination between experiment and theory, resulting from angular anisotropy in the electron yield across the small wedge denoted in panel \textbf{e} of Fig. 1.
We identify an offset of $\sim$2.3\% of the full detector angle.

Figure~3~\textbf{a} shows the vector potential-direction dependent electron yield measured at different x-ray excitation energies (black dots) compared with the simulated yield~(solid line).
The transient revival at $\tau\sim3.5$~fs resulting from electronic coherence in the core-excited state is indicated by the gray arrow and is observed in both experiment and simulation.
This is a quantum beat, occurring at the moment when the quantum phases of the coherently excited $^2\Sigma^-$ and $^2\Delta$ excitations re-align.
This alignment causes constructive interference between the two core-excited states, and an increase in \AM emission rate.
The feature at $\sim$1.3~fs measured at central photon energy 536~eV is likely due to the temporal build-up of the Fano interference between the resonant and direct excitation channels and has been qualitatively reproduced in further simulation.
Analysis of the energetic positions of the Rydberg series converging to the oxygen $K$-edge~\cite{puttner_vibrationally_1999} is not consistent with the interpretation that this modulation is due to further coherent excitation involving Rydberg states.

Figure~3~\textbf{b} shows a zoom-in of the revival feature.
By tuning the central x-ray photon energy away from the center of the $1s\rightarrow2p\pi$ resonance, we are able to suppress the quantum beat in both experiment~(left) and in simulation~(right), demonstrating control over the coherent evolution of the core-excited states. 
The beat is suppressed at higher photon energy due to an increased relative contribution from the direct channel \textit{vs.} the coherently excited resonant decay pathways.
In Figure~3~\textbf{c} we compare our measurement to simulation including (deep red) and excluding (pale red) coherence in the core excited state, for a central x-ray excitation energy of 533~eV.
As expected, it is only possible to reproduce the revival features by including a coherent interaction between the different core-excited electronic states.


\paragraph*{Conclusion}

In conclusion, this work reports the real-time measurement of electronic coherence in the temporal evolution of a core-excited molecule.
Electronic coherence imparts a modulation in the time-dependent emission rate of \AM electrons, driven by an isolated attosecond soft x-ray pulse from a free electron laser.
The \AM emission occurs on a few-femtosecond timescale and we time-resolve it using angular streaking.
Our measurement provides a testbed for exploring the effect of electronic coherence in the photoexcitation dynamics and subsequent photochemical behavior of molecular systems.
The existence of this electronic coherence provides the opportunity to explore inter-atomic site electronic wavepacket coupling, which can reveal interactions between different parts of an extended system~\cite{cederbaum_ultrafast_1999,calegari_ultrafast_2014,kraus_measurement_2015}. 
Measuring this coupling can reveal important information on the system's fundamental physical properties~\cite{cederbaum_correlation_1986,kuleff_ultrafast_2014}.
For example, the spectral makeup of the observed modulations provides rich information on the composition of the excited superposition state.
This opens the possibility to observe the evolution and decay of coherent electronic states in real-time, as they evolve and couple to subsequent nuclear motion in the first stages of a photochemical reaction~\cite{lepine_attosecond_2014,vacher_electron_2017,despre_charge_2018,marciniak_electron_2019}.


\nocite{zholents_method_2005}
\nocite{ammosov_tunnel_1987}
\nocite{storn_differential_1997}
\nocite{bonifacio_collective_1984}

\bibliography{referencesJPC,additionalrefs}
\bibliographystyle{Science}


\paragraph{Acknowledgements}
Support for S. L., J. D., J. P. Mac. and A. M. was provided by DOE, BES Scientific User Facilities Division Field Work Proposal 100317; the contributions from T. D., P. H. B., A. K., A. N., J. T. O., T. J. A. W., A. L. W., and J. P. C. were supported by the U.S. Department of Energy (DOE), Office of Science, Office of Basic Energy Sciences (BES), Chemical Sciences, Geosciences, and Biosciences Division (CSGB); E. G. C. was supported by the DOE Laboratory Directed Research and Development program at SLAC National Accelerator Laboratory, under contract DE-AC02-76SF00515.  
P. R. and M. F. K. acknowledge support by the German Research Foundation \textit{via} KL-1439/10, and the Fellow program of the Max Planck Society.
V. A, J. C. T. B., D. G., J. P. Mar. gratefully acknowledge funding support from UK EPSRC grants No. EP/R019509/1, EP/T006943/1 and No. EP/ I032517/1.
N. B., R. O. and A. C. L. acknowledge the Chemical Sciences, Geosciences and Biosciences Division, Office of Basic Energy Sciences, Office of Science, US Department of Energy, grant no. DE-SC0012376.
C. B. acknowledges the Swiss National Science Foundation and the National Center of Competence in Research – Molecular Ultrafast Science and Technology NCCR - MUST.
L. F. D. and L. F. acknowledge support from NSF Grant No. 1605042 and DOE DE-FG02-04ER15614.
W. H. thanks the German BMBF for funding of the project `SpeAR\_XFEL' under the contract number 05K19PE1.
Use of the Linac Coherent Light Source (LCLS), SLAC National Accelerator Laboratory, is supported by the U.S. Department of Energy, Office of Science, Office of Basic Energy Sciences under Contract No. DE-AC02-76SF00515.


\clearpage

\begin{figure}
\centering
\resizebox{\columnwidth}{!}{\includegraphics{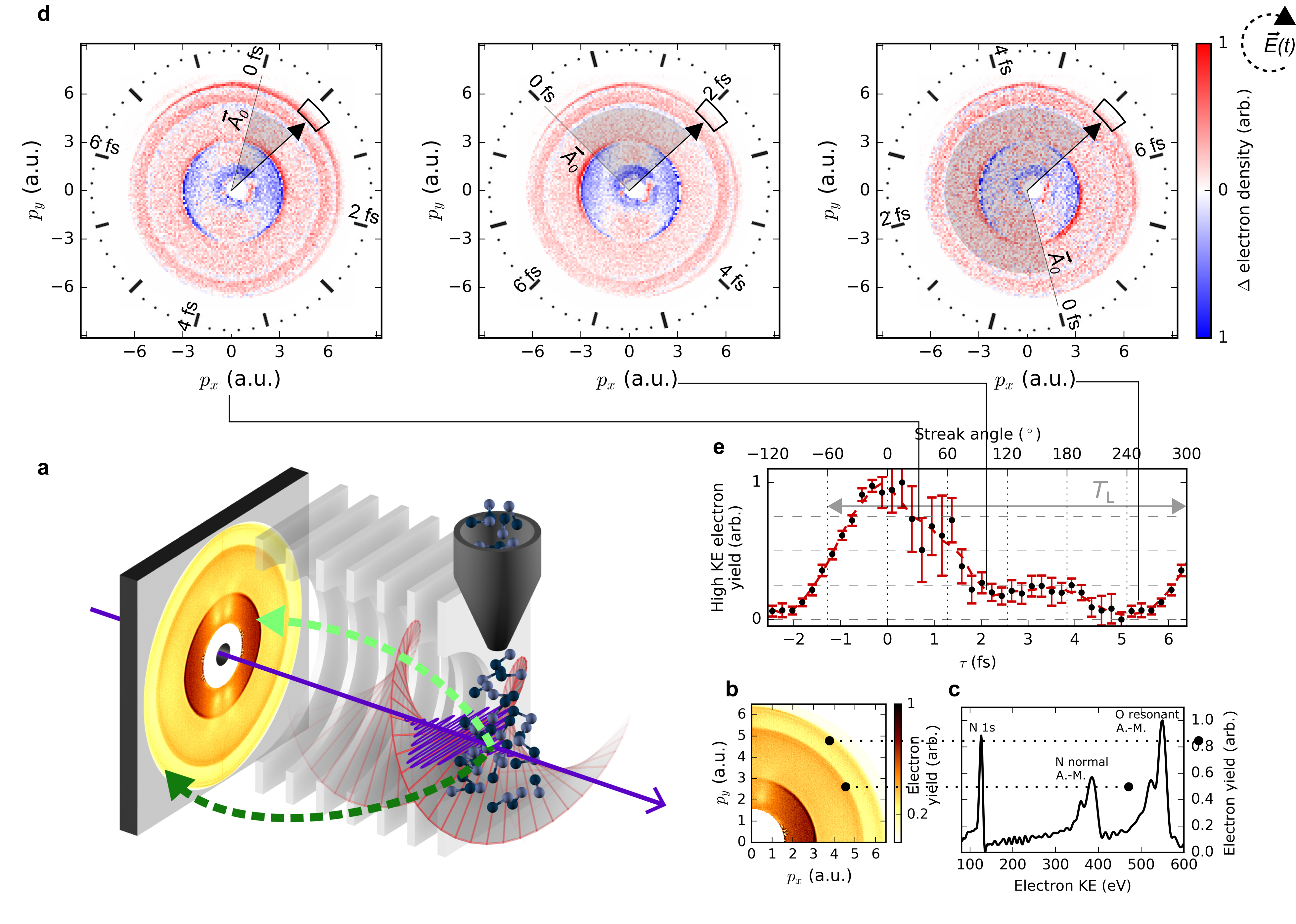}}
\caption{\textbf{a}~NO gas is ionized by an attosecond XFEL pulse at $\sim$530--540 eV in the presence of a 2.3~$\mu$ m circularly polarized streaking field. The resultant photoelectron momentum distribution is measured by a co-axial velocity map imaging spectrometer (c-VMI) \cite{li_co-axial_2018}. The streaking field maps the instantaneous ionization rate onto the measured photoelectron momentum distribution. \textbf{b} Single-color electron momentum spectrum recorded with the c-VMI in the absence of the streaking field. Atomic units denoted here and throughout as `a.u.'. ~\textbf{c}  Applying an inverse Abel transform to this image, we retrieve the electron kinetic energy distribution (`arb.' denotes arbitrary units). \textbf{d} Monitoring the \AM yield on a small (15$^\circ$) region of the detector (black wedge) at different x-ray arrival times produces the time-dependence of \AM emission from core-excited NO, which is shown in black dots in panel~\textbf{e} (dashed red line shows trace with high frequency noise filter applied). The momentum shift of the N $K$-shell photoline (black arrow in panel \textbf{d}) provides a single-shot reference for the direction of $\vec{-A_0}$ at the x-ray pulse time of arrival. $\vec{E}$ shows the direction of rotation of the electric field. The absolute number of measured electrons changes by a factor of 2 between the minimum (normalized to 0) and maximum (normalized to 1).}
\label{fig:expt}
\end{figure}

\begin{figure}
    \centering
    \resizebox{\columnwidth}{!}{\includegraphics{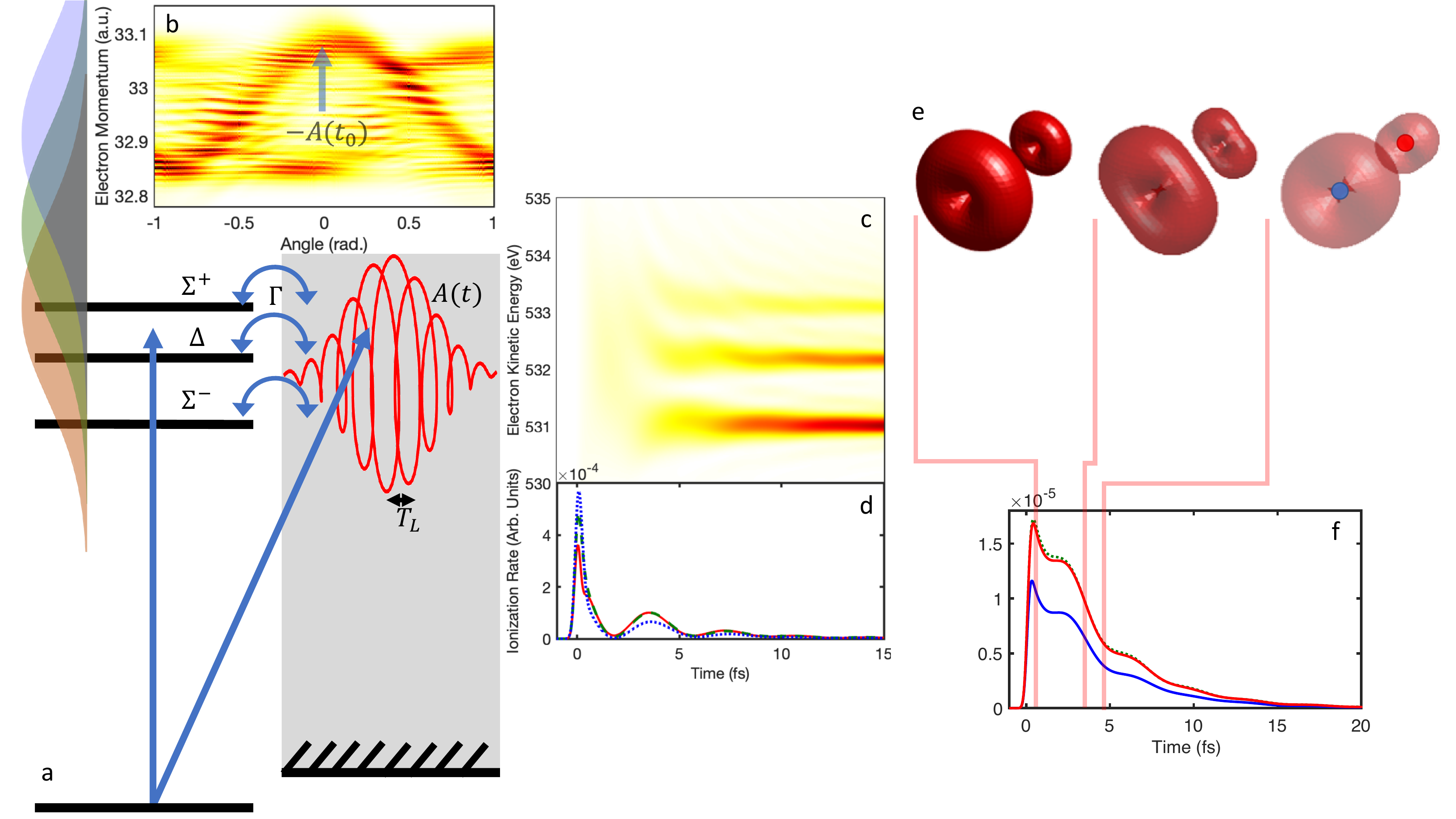}}
    \caption{(a)~schematic showing simple model used for Auger-Meitner emission. The sub-femtosecond x-ray pulse coherently excites three resonances~($^2\Sigma^{+}, ^2\Delta, ^2\Sigma^{-}$). Electrons can also be directly ionized by the x-ray pulse, leading to interfering paths from the ground state to the field-dressed continuum (although the direct ionization pathway is a minor channel \cite{wang_is_2003}). (b)~shows the calculated photoelectron momentum spectrum for $0.5$~fs x-ray pulses centered at $533$~eV photon energy in the presence of a $2.3~\mu$ m laser field. (c)~shows the wavefunction of the continuum electron as a function of time in the absence of the streaking laser field. (d)~shows the ionization rate as a function of time, summed over electron kinetic energy for a central photon energy of $533$~eV~(red),~$534.5$~eV~(green), and~$536$~eV~(blue). (e)~shows the time evolution of the electron density of the bound electronic states. The 3D contour is drawn at 20\% of the maximum electron density and its transparency represents the overall bound-state population, which decays \textit{via} \AM emission. The blue and red dots in the left most panel show the positions of the nitrogen and oxygen atoms, respectively. (f)~shows the total population of the core-excited states as a function of time delay for the same photon energies as in~(d). }
    \label{fig:AugerModel}
\end{figure}

\begin{figure}
\centering
\includegraphics[width=1\textwidth]{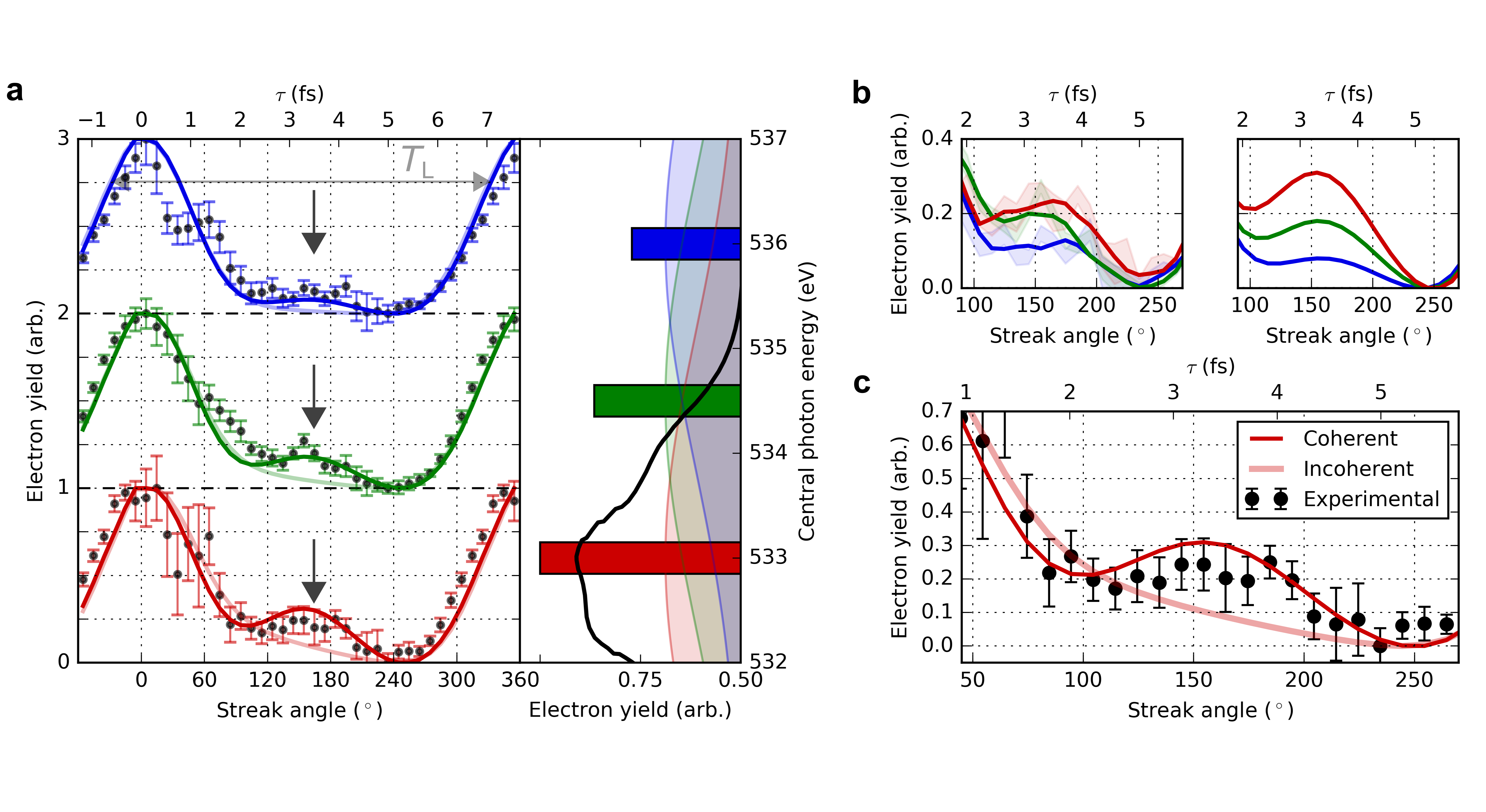}
\caption{\textbf{a}~Direct measurement of time-resolved Auger-Meitner emission from core-excited NO. Left hand panel shows the experimentally measured time-dependent \AM yield as a function of the central XFEL photon energies~(black dots). This is compared with the results of the model shown in Fig.~2~(solid colored lines). Right hand panel shows total electron yield, which decreases as the central photon energy moves away from the center of the 1$s\rightarrow\pi$ resonance (bars). The time-dependent yields change by a factor of 2 between the minimum (normalized to 0) and maximum (normalized to 1) values. The coherent bandwidth of the attosecond XFEL pulse spans $\sim$5 eV, as illustrated by a Gaussian curve of equivalent full width at half maximum at each central photon energy. Black line shows the O$1s\rightarrow\pi$ feature reported in \cite{puttner_vibrationally_1999}, comprising the three electronic states $^2\Sigma^-$, $^2\Delta$ and $^2\Sigma^+$. The revival at $\tau\sim$3.5~fs, marked by the black vertical arrow, is due to the rephasing~(constructive interference) of the \AM emission from the core-excited states~($^2\Sigma$ and $^2\Delta$) populated by the x-ray pulse. The coherent revival is suppressed as the photon energy moves above the 1$s\rightarrow\pi$ resonance and the contribution from the direct photoionization channel increases. The photon energy-dependence of the quantum beat is shown in the zoom-in in panel \textbf{b}, for experiment (left) and simulation (right). \textbf{c}~Comparison between two different models where core-excited states are populated coherently~(deep red) and incoherently~(pale red) at 533~eV central photon energy. The experimental measurement is shown in black dots. Coherent interaction between the core-excited states is required to account for the measured data.}
\label{fig:res}
\end{figure}

\clearpage
\appendix
\renewcommand{\thefigure}{S\arabic{figure}}
\setcounter{figure}{0}
\renewcommand{\theequation}{S\arabic{equation}}
\setcounter{equation}{0}
\newcommand{\duristunable}[0]{(12)}
\newcommand{\wickenhauserattosecond}[0]{(25)}
\newcommand{\haynesclocking}[0]{(9)}
\newcommand{\puttnervibrationally}[0]{(27)}
\newcommand{\glowniatimeresolved}[0]{(20)}
\newcommand{\wangis}[0]{(14)}
\newcommand{\licoaxial}[0]{(13)}
\newcommand{\zholentsmethod}[0]{(38)}
\newcommand{\ammosovtunnel}[0]{(39)}
\newcommand{\storndifferential}[0]{(40)}
\newcommand{\bonifaciocollective}[0]{(41)}

\section{Experimental Methods}

\subsection{XFEL Setup}
Attosecond x-ray pulses are produced using the enhanced self-amplified spontaneous emission (ESASE) technique~\zholentsmethod~employed at the Linac Coherent Light Source in the XLEAP configuration.
The scheme is described in detail in reference \duristunable.
Briefly, coherent undulator radiation generated by the horn of the electron bunch interacts with the bunch as it passes through a long period (35 cm) wiggler.
This creates an energy modulation in the electron bunch, which is turned into a high current spike as the bunch passes through a dispersive magnetic chicane.
This spike is made to lase in the LCLS undulators, producing isolated attosecond soft x-ray pulses with a median measured full-width-at-half-maximum (FWHM) duration of 480 as~\duristunable~at 570 eV photon energy.
The x-ray pulses are focused by a pair of Kirkpatrick-Baez (KB) mirrors to a spot size of $\sim100~\mu$m in the experimental interaction region.

\subsection{Electron Spectrometer}
We used a co-axial velocity map imaging (c-VMI) spectrometer to collect the photoelectrons. 
The spectrometer design is described in detail in reference ~\licoaxial.
The co-axial geometry of the c-VMI refers to the co-axial propagation direction of the x-ray pulse/streaking laser pulse and the photoelectrons, whose momentum distribution is projected by a set of electrostatic plates along the direction of light pulses onto the microchannel plate (MCP) detector.
This preserves the transverse structure of the photoelectron distribution in the plane normal to the direction of propagation of the x-rays. 
This c-VMI spectrometer supports detection of electrons of kinetic energies up to a few hundred electron volts, with an energy resolution of a few percent, depending on the kinetic energy of interest. 
The c-VMI is installed in the experimental end station inside a 10-inch vacuum chamber. 
A 6.4~mm hole at the center of the MCP allows the x-ray pulse and the streaking laser to pass through the detector after the interaction point. 
Since the electrons recorded in this measurement have energies on the order of hundreds of eV, the center hole minimally affects the detection of the electrons of interest.

The gas jet consists of a 2 mm diameter skimmer (Beam Dynamics, Model 2), placed $\sim$120 mm from the interaction region, and a pulsed supersonic gas nozzle (Even-Lavie). The skimmed molecular beam intersects with the x-ray pulse and the streaking laser pulse in the interaction region between the repeller and extractor electrodes inside the c-VMI.

\subsection{Laser Setup}
\begin{figure}
    \centering
    \includegraphics[width=0.5\textwidth]{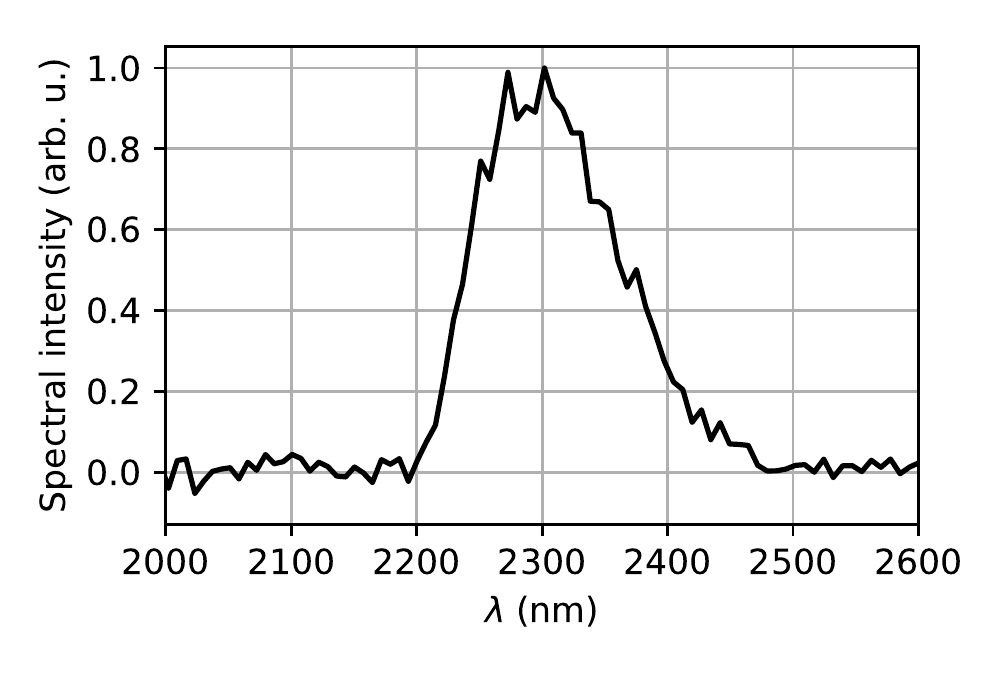}
    \caption{Spectrum of the streaking laser measured immediately before the vacuum chamber.}
    \label{fig:laser_spec}
\end{figure}

The IR streaking laser pulse was generated using a commercial optical parametric amplifier~(OPA, TOPAS-HE). 
The OPA is pumped with $10$~mJ, $\sim50$~fs, $800$~nm pulses from a Ti:sapphire laser system. 
This produces $\sim100~\mu$J idler pulses centered at 2.3~$\mu$m, which are separated from the signal pulse with a dichroic beamsplitter.
After passing through a broad bandwidth quarter-waveplate~(Thorlabs), the beam is focused with an $f=750$~mm lens. 
The focused beam is reflected from a dichroic mirror~(HR: 2400 nm / T: visible) before entering the vacuum system.
The spectrum of the streaking laser pulse measured immediately before it is coupled into the vacuum chamber is shown in Fig.~\ref{fig:laser_spec}.
Inside the vacuum chamber the IR beam is coupled into the interaction region \textit{via} a mirror with a hole, through which the x-rays pass.
This allows the IR beam to co-propagate with the x-ray pulse in the interaction region.

\begin{figure}
    \centering
    \includegraphics[width=0.5\textwidth]{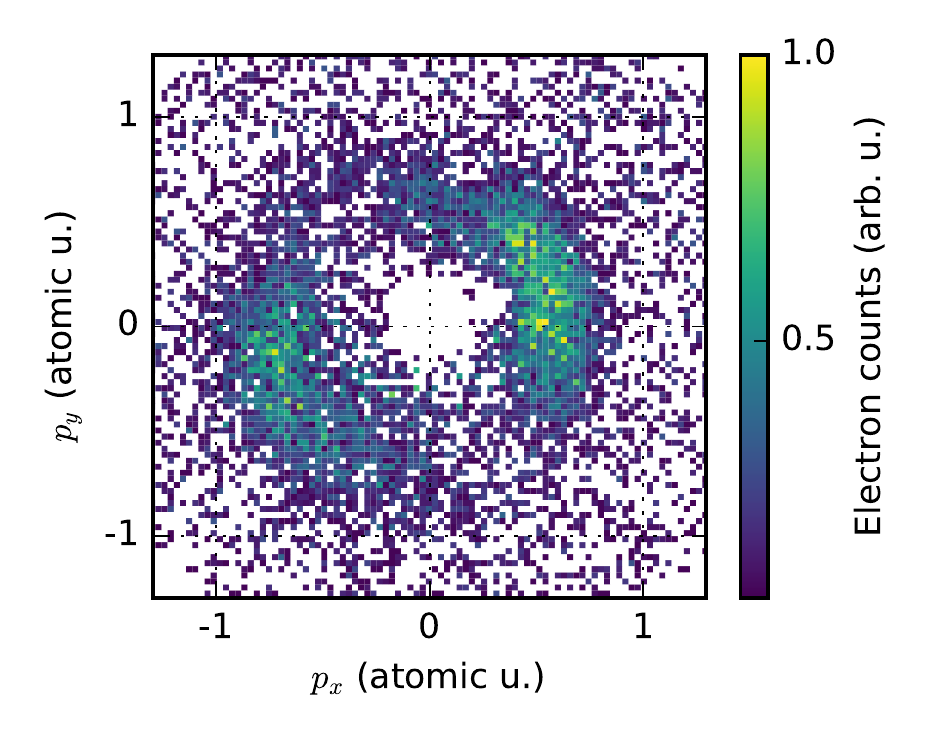}
    \caption{Above-threshold ionization of NO by the 2.3~$\mu$m laser field used for the streaking measurement.}
    \label{fig:ati}
\end{figure}

The intensity of the streaking laser was adjusted with an iris just before the focusing lens. 
The iris size was set such that any ionization from the streaking laser produced $<10$ electrons/shot during the streaking experiments. 
Increasing the IR intensity allowed for characterization of the ellipticity of the streaking laser field. 
Figure~\ref{fig:ati} shows the photoelectron spectrum generated by above-threshold ionization (ATI) of NO molecules using the increased IR intensity and the same quarter-wave plate rotation as in the streaking measurements.
The angular anisotropy of the photoelectron distribution generated from this highly nonlinear process enables accurate determination of the streaking field ellipticity.
By comparing the angular maxima and minima in photoelectron yield with ADK (Ammosov-Delone-Krainov) theory predictions of tunnel ionization rates \ammosovtunnel, we identify an ellipticity of $\geq$0.95.


\section{Data Analysis}
\subsection{X-ray arrival time determination}
Our technique for extracting the time-dependent \AM current relies on a single-shot diagnostic of the streaking laser phase at the time of arrival of the attosecond x-ray pulse.
This is achieved by identifying the momentum shift experienced by the electrons produced from direct ionization of the nitrogen $K$-shell  at each shot.
This self-referencing technique bypasses the requirement for attosecond x-ray/streaking laser temporal stability.
We identify the streaking direction using two different methods and cross-check the extracted value for each shot.
Only shots where the streak direction agrees within 15$^\circ$ are used.

\begin{figure}
    \centering
    \includegraphics[width=\textwidth]{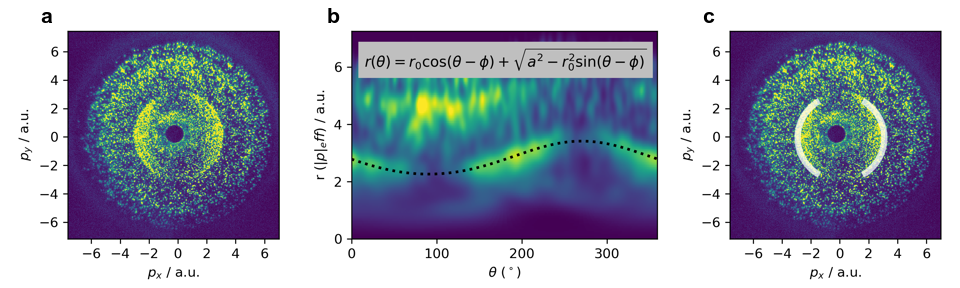}
    \caption{Single-shot identification of x-ray arrival time. \textbf{a} raw c-VMI image of single XFEL shot. The x-ray arrival time is determined by two separate methods: \textbf{b} fitting the radial maxima to the equation for a shifted circle and \textbf{c} maximizing the number of counts that fall within a displaced partial ring (shaded white) representing the angular distribution of the N $K$-shell photoline.}
    \label{fig:streakfind}
\end{figure}

The two methods employed for identifying the single-shot streaking direction of the nitrogen $K$-shell photoline are illustrated in Fig. \ref{fig:streakfind}.
Panel \textbf{a} shows the raw image of the phosphor screen for a single XFEL shot.
The nitrogen $K$-shell photoline is clearly visible in a single shot, at $\left|\vec{p}\right|{\sim}3$ atomic units of momentum.
In the first technique, the raw image of the c-VMI phosphor screen is discriminated with a constant threshold~(all pixels with value less than 90 are set to zero) to remove white noise, and the images are convolved with a Gaussian filter ($\sigma$=25 pixels) and downsized from 1024$\times$1024 to 128$\times$128 pixels.
The resultant filtered and downsized image is rebinned into polar coordinates and the radial maximum of the streaked electron distribution is determined at each detector angle.
The radial distribution of the angle-dependent maxima is then fitted to the equation for a shifted circle in polar coordinates:
\begin{equation}
    r(\theta) = r_0 \cos(\theta - \phi) + \sqrt{a^2 - r_0^2 \sin(\theta - \phi)}
    \label{eqn:shiftedcirc}
\end{equation}
where $r, \theta$ are the polar coordinates and $r_0$ and $\phi$ describe the distance from the origin of the polar coordinate system to the center of a circle of radius $a$.
The origin of the polar coordinates is chosen to be displaced from the center of the unstreaked electron distribution.
The result of this fit is shown in panel \textbf{b} of Fig. \ref{fig:streakfind} for the shot shown in panel \textbf{a}.

The second technique makes use of a genetic optimization algorithm, which maximizes the number of electron counts falling into an open ring as it is translated across the detector image \storndifferential.
The deviation of the center of the ring from the detector origin at this maximal point is determined to be the momentum shift $\vec{A_0}$ induced by the streaking field.
The shape of the ring and its optimized position for this shot are shown in panel \textbf{c} of Fig. \ref{fig:streakfind}.
The results obtained for each image by these two separate analysis procedures are compared, and shots where the methods do not agree within 15$^{\circ}$ are discarded.
The experimental uncertainty associated with our measurement of the streaking laser vector potential at the x-ray pulse arrival time is estimated using a set of measurements taken concurrent with the presented results.
In these measurements we produce two energetically separated photolines at each x-ray shot, from ionization of both the nitrogen and oxygen $K$-shells in nitric oxide.
We choose the x-ray photon energy to be high enough that both photolines experience a momentum shift, according to Eq. 1 of the main text, in the direction of the streaking laser vector potential at the x-ray pulse time of arrival.
We independently determine the vector potential at the time of ionization using each of the two photolines, and calculate the distribution of the difference in determined vector potential between each photoline.
Assuming the error to be evenly distributed between the two photolines, we identify an error distribution of $\sigma$=30~$^\circ$ for single shot vector potential determination.
Therefore, a Gaussian convolution ($\sigma$=30$^\circ$) is applied to the simulated time trace as described below, to account for the experimental error in x-ray arrival time determination.

The electron yield on the small region of the detector illustrated in panel~\textbf{d} of Fig.~1 in the main text is monitored as a function of x-ray arrival time.
The XFEL shots are binned into 36 different bins according to the x-ray arrival time.
The electron yield in the detector region is determined at each shot using a 2-D hit-finder routine which determines local maxima on the filtered full image.
The electron count in this region of the detector is sparse enough to enable counting single electron hits.
The hit-finder is employed to limit the effect of changes in detector gain on the measured electron yield.
The 15$^\circ$ region of the detector is divided into individual 1$^\circ$ regions following polar rebinning of the photoelectron momentum distribution.
The correlation between the electron yield at each of the 15 different regions on the detector was investigated and found to be negligible.
The error bars presented in the main text represent twice the standard error of the electron yield recorded across each of these 1$^\circ$ regions.

\section{Modeling the core-excited resonance}
The model used to describe the data in figures~2 and~3 in the main text 
is an application of the method published by Wickenhauser~\textit{et al.} in Ref.~\wickenhauserattosecond. 
The full Hamiltonian for the system is
\begin{equation}
    H(t) = H_0 + V + H_X(t) + H_L(t),
\end{equation}
where $H_0$ is the atomic Hamiltonian describing single electronic configurations, $V$ describes the interaction between the single configurations of the bound and continuum states, and $H_X(t)$ and $H_L(t)$ describe the x-ray and streaking laser interactions, respectively, within the dipole approximation.
Wickenhauser~\textit{et al.} describe calculation of the time- and momentum-differential ionization probability $P(\vec{p},t)=\left|\left<\vec{p}\left|\psi(t)\right.\right>\right|^2$, where $\psi(t)$ is the solution to the time-dependent Schrodinger equation,
\begin{eqnarray}
    \left|\psi(t)\right\rangle &=& -i \int_{t_0}^t dt^{\prime}U(t,t^{\prime}) H_X(t^{\prime})\left|g\right\rangle, \\
    &&U(t_2,t_1) = \exp{\left[-i\int_{t_1}^{t_2} dt \left[H_0 + V + H_L(t)\right] \right]} \nonumber
\end{eqnarray}
to first order in the x-ray matter interaction \wickenhauserattosecond. 
The momentum distribution measured in the experiment is then given by the asymptotic limit $P(\vec{p})=P(\vec{p},t\rightarrow\infty)$.

The interaction with the IR field is highly non-perturbative, and therefore the solution to the time-dependent Schrodinger equation is expanded within the strong-field approximation~(SFA), 
\begin{subequations}
    \label{eqn:supp:IonAmp}
    \begin{align}
    \left\langle \vec{p}\left| \psi(t)\right\rangle\right. &= -i \int_{t_0}^t d t^{\prime} e^{i\Phi_V(t^{\prime},t)} \left\langle \vec{p}\left|H_X(t^{\prime})\left|g\right.\right.\right\rangle \label{eqn:supp:IonAmp_a} \\
    &-\sum_k \sum_r \int_{t_0}^t d t^{\prime} \int_{t^{\prime}}^t dt^{\prime\prime} e^{i\Phi_V(t^{\prime\prime},t)} V_{E(t^{\prime\prime})r}\left\langle r\left|U_F(t^{\prime\prime},t^{\prime})\left|k\right.\right.\right\rangle\left\langle k\left|H_X(t^{\prime})\left|g\right.\right.\right\rangle \label{eqn:supp:IonAmp_b}\\
    &-\sum_k \int_{t_0}^t d t^{\prime} \int_{t^{\prime}}^t d t^{\prime\prime} \int d\vec{p}^{\prime} e^{i\Phi_V(t^{\prime},t)} V_{E(t^{\prime\prime})k}\left\langle k\left|U_F(t^{\prime\prime},t^{\prime})\left|\vec{p}^{\prime}\right.\right.\right\rangle \left\langle \vec{p}^{\prime}\left|H_X(t^{\prime})\left|g\right.\right.\right\rangle \label{eqn:supp:IonAmp_c},
    \end{align}
\end{subequations}
where
\begin{equation}
    \Phi_V(t_0,t) = \int_{t_0}^t dt^{\prime}\frac{\left[\vec{p} - A(t^{\prime})\right]^2}{2} + I_p
\end{equation}
is the Volkov phase, 
\begin{equation}
    E(t) = \frac{1}{2}\left[\vec{p} - \vec{A}(t)\right]^2,
\end{equation}
is the energy of the electron in the continuum as a function of time, and
\begin{eqnarray}
        U_F(t_2,t_1) &=& \exp{\left[-i\int_{t_1}^{t_2} dt H_F \right]}, \\
        &&H_F = \left[ \begin{array}{ccccc}
             E_{k_1} & 0 & \cdots &\Gamma_{k_1p} & \cdots \\
             0 & E_{k_2} & 0 &\Gamma_{k_2p} & \cdots \\
             \vdots & 0 & \ddots & \Gamma_{kp} & \cdots \\
             \Gamma_{k_1p} & \Gamma_{k_2p} & \Gamma_{kp} & \nicefrac{p^2}{2} & 0\\
             \vdots & \vdots & \vdots & 0 & \ddots
        \end{array} \right] \nonumber
    \label{eqn:supp:HF}
\end{eqnarray}
is the propagator for the Fano states. 
The matrix element $\left\langle r\left|U_F(t,t_0)\left|k\right.\right.\right\rangle$ describes the survival probability, i.e. the probability of finding the system in state $r$ at time $t$ assuming that the system was in state $k$ at time $t_0$.
The first line in Eqn.~\ref{eqn:supp:IonAmp} represents the direct ionization of the ground state to the dressed continuum. 
The second line~(\ref{eqn:supp:IonAmp_b}) describes the process where the system is initially excited to a bound state~($k$) and transitions to the dressed continuum after some time $t^{\prime\prime}-t^{\prime}$.
During the time interval $t^{\prime\prime}-t^{\prime}$ the system can transition between the bound and continuum states many times, but the final transition occurs at time $t^{\prime\prime}$.
Finally, line~\ref{eqn:supp:IonAmp_c} describes a process where the system is initially excited to the continuum at energy $E=\nicefrac{p^{\prime2}}{2}$ at time $t^{\prime}$, but transitions back to the bound state~($k$) in the interval $t^{\prime\prime}-t^{\prime}$, before making a final transition to the dressed continuum at energy $E=\nicefrac{p^2}{2}$.
The sums $k$ and $r$ run over all of the core-excited states~($\Sigma^+$,$\Delta$,$\Sigma^-$).
Assuming the configuration interaction matrix elements $V_{Ek}$ are energy-independent~(i.e. $V_{Ek}=V_{k}=\Gamma$ for all $k$), Wickenhauser~\textit{et al.} showed that this expression can be simplified by analytical integration,
\begin{eqnarray}
    \label{eqn:supp:IonAmpSimp}
    \left\langle E\left| \psi(t)\right\rangle\right. &=& -\int_{t_0}^t d t^{\prime} e^{i\Phi_V(t^{\prime},t)} \\
    &&\times \left[ i \mathcal{E}_X(t^{\prime}) \left\langle E\left|d\left|g\right.\right.\right\rangle  + \sum_{k,r} \left(1+\frac{i}{q_k}\right)\int_{t_0}^{t^{\prime}} dt^{\prime\prime} \mathcal{E}_X(t^{\prime\prime}) \Gamma \left\langle r\left|U_F(t^{\prime\prime},t^{\prime})\left|k\right.\right.\right\rangle\left\langle k\left|d\left|g\right.\right.\right\rangle
    \right]\nonumber,
\end{eqnarray}
where
\begin{equation}
    q_k = \frac{\left\langle k\left|d\left|g\right.\right.\right\rangle}{\pi\Gamma\left\langle E\left|d\left|g\right.\right.\right\rangle}
\end{equation}
is the Fano parameter determined by the relative amplitudes between the dipole matrix elements for direct ionization to the continuum~($\left\langle E|d|g\right\rangle$) \textit{vs} resonant excitation~($\left\langle k|d|g\right\rangle$) \wickenhauserattosecond.
This quantum mechanical treatment of the resonant Auger-Meitner  process is consistent with the model used by Haynes~\textit{et al.} to describe the normal \AM decay process~\haynesclocking.

\subsection{Including Incoherent Effects }
\begin{figure}
    \centering
    \includegraphics[width=0.6\textwidth]{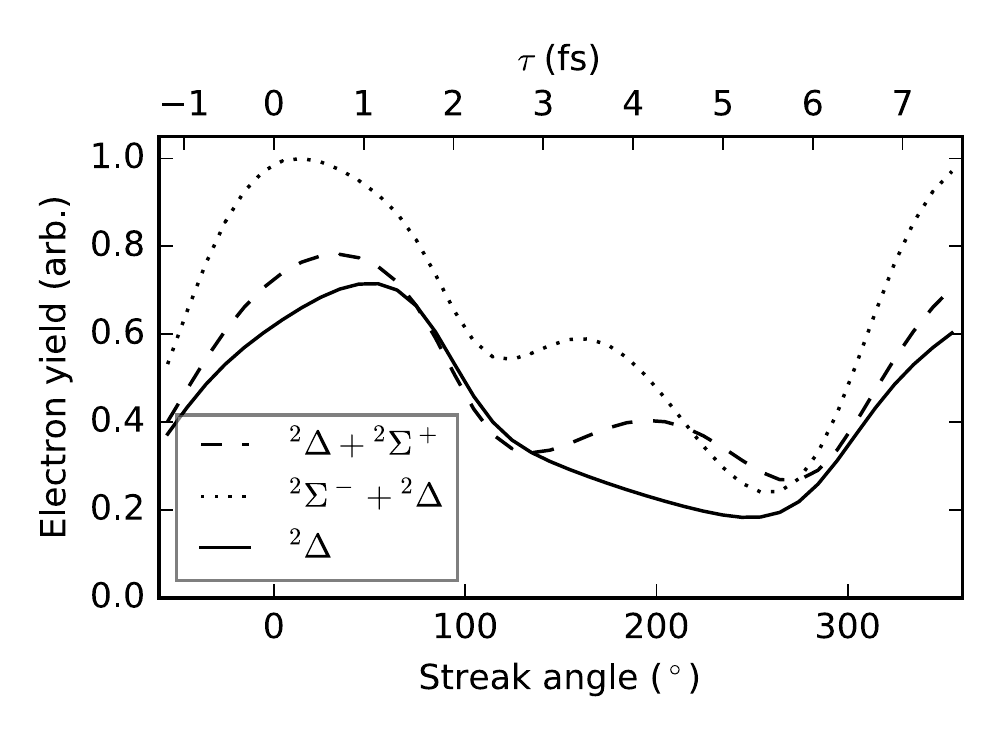}
    \caption{Time-dependent \AM emission rates for three sets of states at central photon energy 533 eV. The three traces are summed incoherently for comparison with experiment.}
    \label{fig:diffstates}
\end{figure}

In this work, Eqn.~\ref{eqn:supp:IonAmpSimp} is numerically integrated to find $\psi(E,t\rightarrow\infty)$.
We choose values of $q_k$ to most closely match the measured absorption spectra from Ref.~\puttnervibrationally, $\vec{q}=[10,8.16, 5.77]$.
There are a number of effects that are not fully described by the model, for example the angular distribution of \AM emission for the different core-excited states.
In the analysis we integrate a 15$^{\circ}$ section of the detector, which leads to a partial loss of coherence because the core-excited states couple to continua of different symmetry~\wangis.
The correct description of this measurement involves tracing the full density matrix over the direction of the outgoing electron and then calculating the observed electron yield.
Instead of re-deriving Eqn.~\ref{eqn:supp:IonAmpSimp} to describe the time evolution of the density matrix, we account for this effect by calculating the signal for each individual state in isolation and for different pairs of states.
The results for each calculation at 533~eV central energy is shown in Fig.~\ref{fig:diffstates}.
Each of these signals is then added together incoherently with coefficients optimized for consistency with the measured data.

\subsection{Accounting for Experimental Parameters in Comparison to the Model}
Within the model described above we calculate the asymptotic momentum distribution of the photoelectrons.
This is the observable accessed by velocity map imaging, allowing direct comparison of measurement and simulation.
We calculate the momentum distribution for 32 x-ray arrival times, evenly spaced within the period $T_L$ of the streaking field.

\begin{figure}
    \centering
    \includegraphics[width=0.6\textwidth]{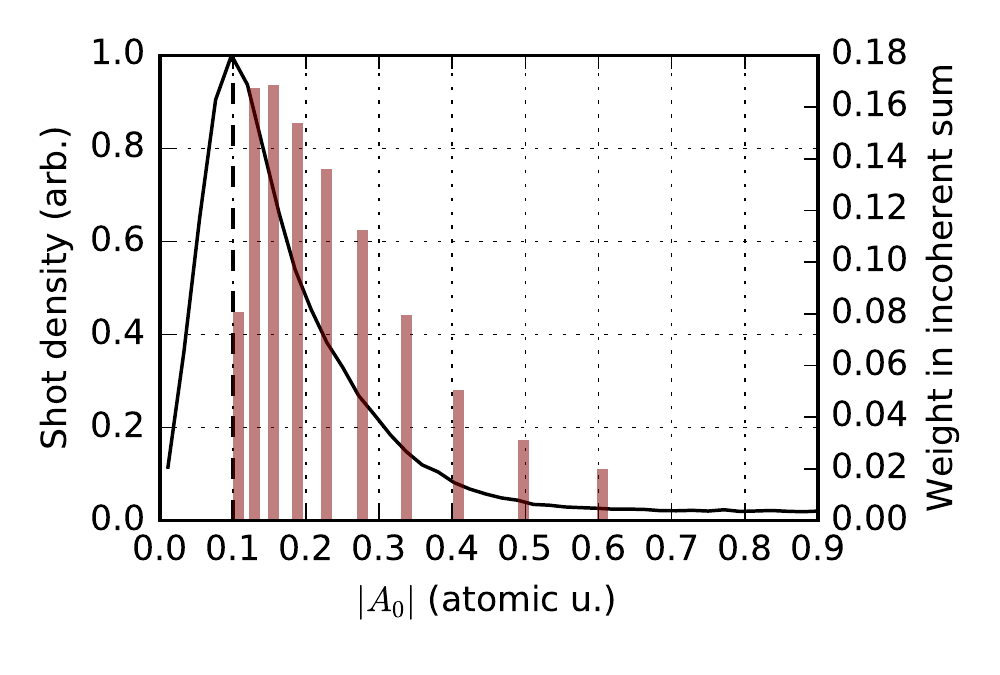}
    \caption{Measured distribution of $|\vec{A}|$ according to shift of nitrogen $K$-shell photoline (black line) and corresponding weight attributed to simulation at given value of $|\vec{A}|$ (red bars).}
    \label{fig:momdist}
\end{figure}

Due to temporal jitter in the absolute arrival time of the x-ray pulse, the synchronization between the streaking laser and x-ray pulse is $\sim500$~fs~\glowniatimeresolved.
This is highly comparable with the streaking laser pulse duration ($\sim$100~fs), resulting in significant shot-to-shot variation in the magnitude $|\vec{A_0}|$ of the streaking field vector potential at the arrival time of the x-ray pulse.
Our single-shot diagnostic for determining the momentum shift experienced by the N~$K$-shell photoelectrons also allows us to access the single-shot value of $|\vec{A_0}|$ at the x-ray time of arrival.
The measured distribution in the ponderomotive shift $U_p$ due to the streaking field is shown in Fig. \ref{fig:momdist}.
To account for this distribution we perform simulations with ten different values of $|\vec{A_0}|$ and incoherently sum the resultant photoelectron momentum distributions, weighted according to the measured distribution of $|\vec{A}|$.
Experimental shots with measured $|\vec{A_0}|<0.1$~atomic units (indicated by the vertical dashed line in Fig. \ref{fig:momdist}) are excluded from the analysis because the error in determining $\frac{\vec{A_0}}{|\vec{A_0}|}$ increases at small $|\vec{A_0}|$.

The single-shot photon energy diagnostic available for this measurement was the relativistic energy of the electron beam following acceleration.
Shot-to-shot variation in photon energy is inherent to SASE operation of an XFEL, but in standard operation the single shot photon energy can be determined to within better than 1~eV using this electron beam energy and the FEL resonance formula~\bonifaciocollective:
\begin{equation}
    \lambda_r = \lambda_u \frac{1+\frac{K_u^2}{2}}{2\gamma^2}.
    \label{eqn:feleqn}
\end{equation}
Here $\gamma$ is the Lorentz factor of the electron beam, $\lambda_u$ and $K_u$ are the undulator period and strength parameters, and $\lambda_r$ is the wavelength of the XFEL radiation.
Although the XFEL pulse builds up from spontaneous emission which initiates at a different, random position in the lasing electron bunch at each shot, in standard SASE operation the bandwidth across which spontaneous emission can occur in the electron bunch is small enough that the electron beam energy can be reliably employed as a single-shot photon energy diagnostic.
In ESASE operation the much larger bandwidth of the lasing spike~\zholentsmethod~reduces the single-shot predictive power of the electron beam energy for x-ray photon energy and the FWHM of the error distribution between measured central photon energy and that given by Eq.~\ref{eqn:feleqn} has been measured at $\sim$3~eV \duristunable.

We account for this by simulating the x-ray/streaking interaction at ten evenly spaced central photon energies between 531.1--537.85~eV and calculating the incoherent weighted sum across the different asymptotic photoelectron momentum distributions.
The distribution at each simulated photon energy is weighted according to the experimentally determined photon energy distribution.
This is calculated by convolving the photon energy distribution measured according to electron beam energy, with a Gaussian curve of 3~eV FWHM.

The linear scaling between photoelectron momentum and radial position on the cVMI detector acquires additional nonlinear terms at high ($\geq$400 eV) electron energy.
Simulation of high energy electron trajectories through the electrostatic lens stack of the c-VMI spectrometer does not indicate this to be a result of different VMI focusing conditions at high kinetic energy~\licoaxial.
Simulation of the optical setup used to image the phosphor detector indicates the probable cause of this nonlinear behavior is aberration at the outer edge of the lens used to image the phosphor screen onto the CCD detector (in VMI operation, higher energy electrons are focused to higher radii and therefore closer to the edge of the detector). 
This results in a degradation of electron kinetic energy resolution at the position of the resonant oxygen \AM emission in NO, which we estimate to be $\nicefrac{\Delta\mathbf{E}}{\mathbf{E}}\sim5\%$ by comparison of our measured \AM spectrum with previous high-resolution experiments.
We account for this effect by applying a Gaussian blur of 5\% $\nicefrac{\Delta\mathbf{E}}{\mathbf{E}}$ to the radial dimension of the simulated c-VMI measurement.

\begin{figure}
    \centering
    \includegraphics[width=0.5\textwidth]{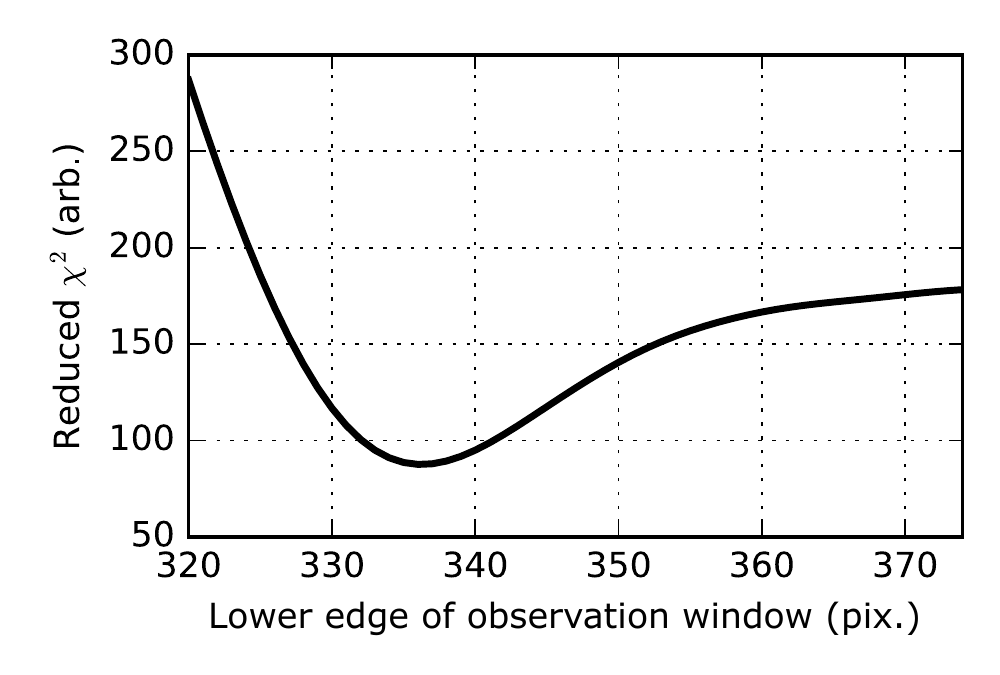}
    \caption{Dependence of reduced $\chi^2$ best fit parameter for fit between experimental and simulated \AM time-dependent trace, on lower edge of electron observation window on detector.}
    \label{fig:lolim}
\end{figure}

Further work is required to fully characterize the nonlinear momentum scaling at high electron energies, but in this work we account for the nonlinearity by incorporating the position of the lower edge of the high energy electron observation window (panel \textbf{d} of Fig.~1 of main text) as a free parameter in the fitting procedure described in the main text.
This lower edge is set by a pixel coordinate on the CCD detector in experiment, but the nonlinear behavior described above complicates the absolute mapping of this position to photoelectron momentum.
No electron counts are recorded above the upper edge of the observation window in experiment, so the simulated electron distribution is integrated to $\vec{p}=\infty$.
Figure \ref{fig:lolim} shows the deviation between experimental and best-fit simulated time-dependent electron yields summed across each of the three central photon energies shown in Fig. 3 in the main text, as the lower edge of the observation window is scanned in simulation.
This curve shows a clear minimum, identifying the position on the simulated detector which corresponds to the lower edge of the experimental observation window.

\subsection{Dependence on x-ray pulse properties}
\begin{figure}
    \centering
    \includegraphics[width=0.6\textwidth]{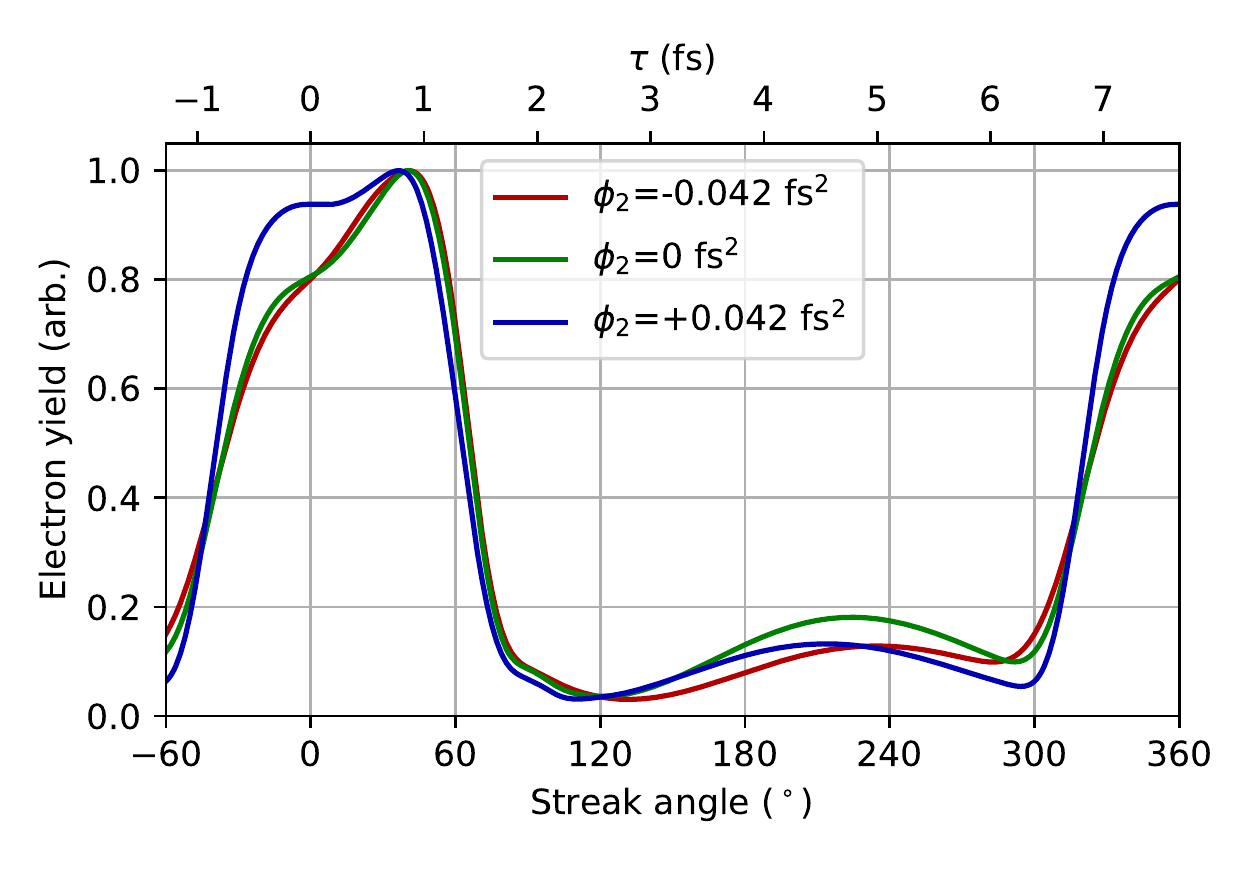}
    \caption{Modification of the simulated time-dependent \AM electron yield resulting from changes to the spectral phase of the exciting x-ray pulse. The quadratic term of the spectral phase $\phi_2$ is adjusted and the time-dependent \AM yield, measured by angular streaking, is simulated according to the model described in section 3.}
    \label{fig:pulsepars}
\end{figure}
The nature of the core-excited electron wavepacket is sensitive to the properties of the x-ray pulse which produces it.
We have found in simulation that our observable of the time-dependent \AM electron yield is sensitive to these changes.
Agreement between simulation and experiment relies on using the correct x-ray parameters in simulation.
Figure \ref{fig:pulsepars} shows the simulated time-dependent high-energy electron yield produced by coherent excitation of NO by x-ray pulses centered at 534.48~eV, calculated for a single vector potential and $\gamma$=0.146~eV.
The linear chirp of the x-ray pulse is adjusted in simulation and this change manifests as a modification of the time-dependent electron yield.
Due to the chirp-taper matching condition of the undulator configuration used in their production \duristunable, the attosecond x-ray pulses employed in our measurement have a chirp of $\sim$-0.042 fs$^2$.
The x-ray pulses employed in simulation are adjusted to match this value.


\end{document}